\documentclass[12pt]{article}

\usepackage{scicite}

\usepackage{times}

\usepackage{color} 
\usepackage{xcolor} 
\usepackage{nicefrac}
\usepackage{amsmath}
\usepackage{amssymb}
\usepackage[hypertexnames=false]{hyperref}
\usepackage{graphicx}
\usepackage[separate-uncertainty = true]{siunitx} 
\usepackage{caption}
\usepackage{sectsty} 
\usepackage{float} 

\usepackage{booktabs}

\usepackage{braket}

\usepackage{lipsum} 

\usepackage{xr}
\newcommand{\iset}{I_{\mathrm{set}}}
\newcommand{\vbias}{V_{\mathrm{bias}}}
\newcommand{\vmod}{V_{\mathrm{mod}}}
\newcommand{\vstab}{V_{\mathrm{stab}}}
\newcommand{\istab}{I_{\mathrm{stab}}}
\newcommand{\didv}{\mathrm{d}I/\mathrm{d}V}

\definecolor{stateblue}{rgb}{0.18,0.5,.75} 
\definecolor{statered}{rgb}{0.88,0.29,.29}

\newcommand{\didu}[0]{$\mathrm{d}I / \mathrm{d}V$}

\definecolor{magenta}{rgb}{1, 0, 1}

\definecolor{purple}{rgb}{0.86, 0.44, 0.84}

\definecolor{orange}{rgb}{1, 0.7, 0}

\newenvironment{sciabstract}{%
\begin{quote} \bf}
{\end{quote}}

\title{Non-local detection of coherent Yu-Shiba-Rusinov quantum projections}

\author
{Khai Ton That$^{1,\dagger}$, Chang Xu$^{2,\dagger}$, Ioannis Ioannidis$^{3,4,\dagger}$, Lucas Schneider$^{1}$,\\
Thore Posske$^{3,4}$, Roland Wiesendanger$^1$, Dirk K. Morr$^2$, and Jens Wiebe$^{1,\ast}$, 
\\
\normalsize{$^{1}$Department of Physics, University of Hamburg, Jungiusstrasse 9A, 20355 Hamburg, Germany.}\\
\normalsize{$^{2}$University of Illinois at Chicago, Chicago, Illinois 60607, USA.}\\
\normalsize{$^{3}$I. Institute for Theoretical Physics, University of Hamburg, D-22607 Hamburg, Germany.}\\
\normalsize{$^{4}$Centre for Ultrafast Imaging, Luruper Chaussee 149, D-22761 Hamburg, Germany.}\\
\\
\normalsize{$^\dagger$These authors contributed equally to this work.}\\
\normalsize{$^\ast$To whom correspondence should be addressed; E-mail:  jwiebe@physnet.uni-hamburg.de.}\\
}

\date{} 

\begin{document} 


\baselineskip24pt


\maketitle

\begin{sciabstract}
Probing spatially confined quantum states from afar – a long-sought goal to minimize external interference – has been proposed to be achievable in condensed matter systems via coherent projection. The latter can be tailored by sculpturing the eigenstates of the electron sea that surrounds the quantum state using atom-by-atom built cages, so-called quantum corrals. However, assuring the coherent nature of the projection, and manipulating its quantum composition, has remained an elusive goal. Here, we experimentally realize the coherent projection of a magnetic impurity-induced, Yu – Shiba – Rusinov quantum state using the eigenmodes of corrals on the surface of a superconductor, which enables us to manipulate the particle-hole composition of the projected state by tuning corral eigenmodes through the Fermi energy. Our results demonstrate a controlled non-local method for the detection of magnet-superconductor hybrid quantum states.
\end{sciabstract}


\section*{Introduction}

Atomic manipulation techniques have provided the unprecedented ability to build atomically precise 
structures of adatoms \cite{Eigler1990, Choi_RevModPhys2019,Khajetoorians_Review2019}
or molecules \cite{Piquero-Zulaica2022} 
on the surface of solid state materials, giving rise to intriguing quantum phenomena ranging from the creation of 
standing matter waves in quantum corrals~\cite{Crommie1993}, and the design of quantum drums~\cite{Moon2008} to the engineering of artificial electronic structures in molecular graphene\cite{Gomes2012}.
Crucial for the emergence of all of these
phenomena is the existence of vertically confined, two-dimensional 
Shockley surface-states~\cite{Shockley1939,Gartland1975,Heimann1977} on $(111)$ surfaces of noble metals that  
allow for the multiple and coherent scattering of electrons off the adatom structure. The resulting constructive 
interference was argued to be the basis for the spatial projection of the many-body Kondo resonance between the foci of an elliptical quantum corral~\cite{Manoharan:2000}. This
effect -- dubbed quantum mirage -- was subsequently theoretically shown to be tied to the realization of the "particle-in-a-box" problem in quantum corrals with tight walls,
exhibiting electronic states -- corral eigenmodes -- with well-defined energies and two-dimensional spatial patterns~\cite{Fiete2001,Porras2001,Agam2001,Weissmann2001,Aligia2001,Hallberg2002,Lobos2003,Schmid2003}. Indeed, it was theoretically predicted that these eigenmodes could also enable the projection of other atomic-scale quantum phenomena, such as the impurity-induced Yu-Shiba-Rusinov (YSR) state in a superconductor (SC)~\cite{Morr2004}, resonant impurity states in topological insulators~\cite{Fu2011}, or 
vibrational~\cite{Gadzuk2003} and spin excitations~\cite{Fransson2012} of atoms, 
opening a unique opportunity to non-locally probe quantum states "at-a-distance" via their projected images. 
This remote detection, in turn, requires that the projected image reflects and maintains the fundamental quantum nature of the original state. How the quantum nature of the projection can be experimentally detected, has remained an intriguing question.

\begin{figure}[H]
\centering	
 \includegraphics[width=1\textwidth]{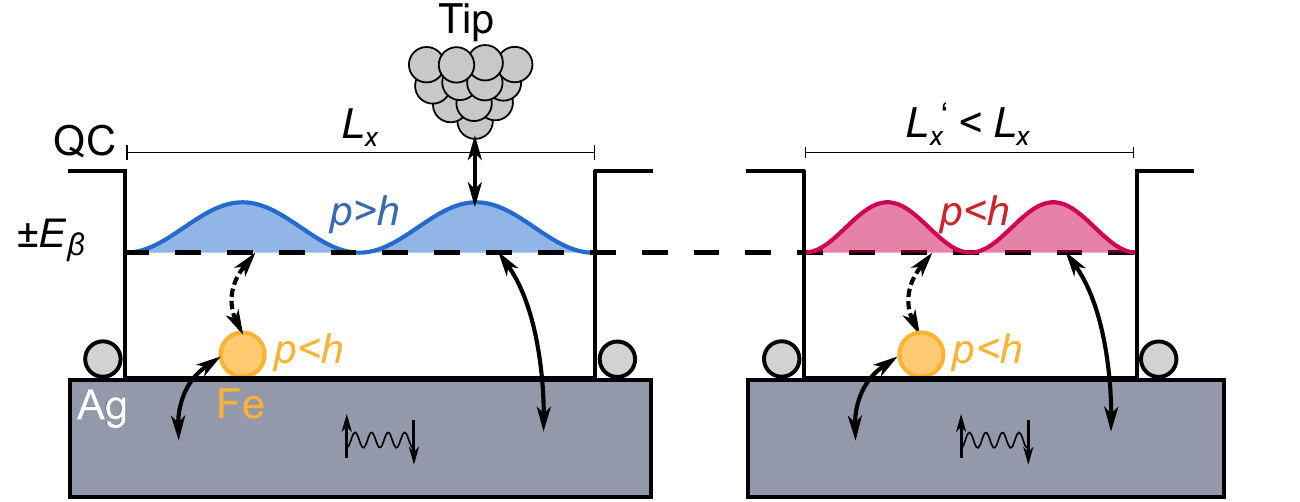} 
 	\caption{\label{fig0} \textbf{Non-local detection and tuning of $p$-$h$ composition in a YSR quantum projection.} Sketch of the YSR quantum projection (wave at energy $\pm E_{\beta}$) in a quantum corral (QC) of Ag atoms (gray spheres) on a proximity-SC Ag island (gray box), induced by the YSR state of an Fe atom (orange sphere). The quantum projection is dominated by a weak, indirect coupling between the YSR state and the quantum corral eigenmode via the proximitized Ag island (continuous lines with arrows), while the direct coupling is negligible (dashed line with arrows). The particle ($p$)-hole ($h$) composition of the projection can be tuned between inverted (left panel, $p>h$) with respect to that of the native YSR state ($p<h$) and non-inverted (right panel, $p<h$) by adjusting the quantum corral geometry, e.g. its lengths ($L_x$ and $L_x^\prime$). The YSR state is non-locally detected using the STM tip.}
 \end{figure}

Here, we demonstrate the quantum nature of a mirage by considering a magnetic impurity-induced and strongly localized YSR state in an $s$-wave SC~\cite{Balatsky2006,Heinrich2018}.
In particular, we show the formation of a hybrid quantum state between the YSR state and a quantum corral eigenmode, with a well-defined phase relation between the spatial oscillations of its particle- ($p$) and hole- ($h$) like components, similar to that found in long-range coherent YSR states~\cite{Balatsky2006,kim2020long,Menard2015}. In the following, we refer to this hybrid quantum state, which was theoretically predicted 20 years ago~\cite{Morr2004}, as coherent quantum projection of the YSR state. To realize the latter, we place single magnetic Fe atoms in Ag-atom based quantum corrals assembled on the $(111)$ surface of thin proximity-SC Ag islands grown on Nb$(110)$, making use of their energetically and spatially well-defined eigenmodes (see Fig.~\ref{fig0}). By coupling to the bulk Ag states, the Fe atom induces a YSR state at energy $\pm E_{\beta}$
deep inside the SC gap. This ensures that its projected spectroscopic image is clearly separated from the de Gennes -- Saint James coherence states of the substrate~\cite{Ortuzar2023} as well as the Machida -- Shibata states (MSSs) which enter the gap whenever a corral eigenmode is close to $E_{\rm F}$~\cite{Schneider2023}.
We demonstrate that an image of this YSR state can be coherently projected by the quantum corral eigenmodes over length scales up to 20 times larger than its localization length, where it is non-locally detected with the STM tip in a minimally perturbative fashion. Over these distances, the $p$- and $h$-like components of the YSR projection maintain their spatial phase shift which is inherited from the MSS, reflecting one of the hallmarks of the long-range coherent YSR quantum state
~\cite{Balatsky2006,kim2020long,Menard2015}.
This projection occurs whenever the Fe adatom is located close to a maximum in the corral eigenmode's wave-function, even when this maximum is not located close to a focus of an elliptical quantum corral. 
By adjusting the corral length $L_x$ and thereby tuning the corral's eigenmodes through $E_{\rm F}$, we can invert the $p$-$h$ composition of the projected YSR state (Fig.~\ref{fig0}), thus essentially manipulating the extent of the (local) $p$-$h$ mixing in the SC state. Our results demonstrate for the first time that the long-range coherent nature of a quantum state can be maintained in a projected image using quantum corrals,
thus opening the door for the non-local probing of more fragile magnet-superconductor hybrid quantum states.

\section*{Native YSR states of Fe-atoms}

\begin{figure}[H]
 	\centering
     \includegraphics[width=0.5\textwidth]{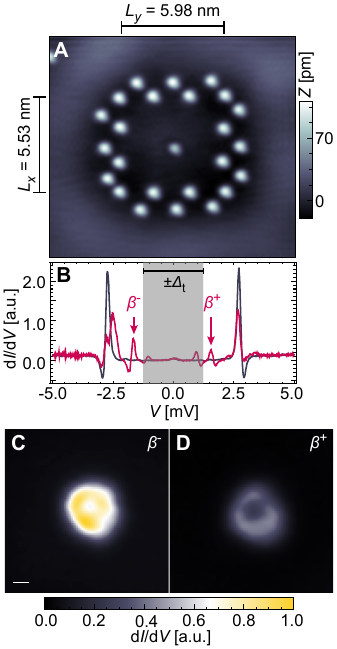}
      \caption{\label{fig1}\textbf{Native YSR states of the Fe atom.} (\textbf{A}) Constant-current STM image of a single Fe atom in the center of a Ag corral of dimensions $L_x =\SI{5.53}{\nano \metre}$ and $L_y =\SI{5.98}{\nano \metre}$ ($\vbias=\SI{5}{\milli \volt}$,
      $\iset =\SI{1}{\nano \ampere}$). $L_x$ and $L_y$ are defined as the distances between the inner rows of Ag atoms. 
      (\textbf{B}) $\didv$ spectra taken on the Fe-atom (red) shown in (A) and a substrate region (black) inside an even smaller corral without Fe atom (fig.~S2, $\vstab=\SI{5}{\milli \volt}$, $\istab=\SI{1}{\nano \ampere}$, $\vmod = \SI{50}{\micro \volt}$
      for the Fe-atom and $\vmod = \SI{20}{\micro \volt}$
      for the substrate).
      (\textbf{C} and \textbf{D}) Constant-contour $\didv$ maps of the Fe atom shown in (A) acquired at the bias voltage $\vbias=$ $\SI{-1.61}{\milli \volt}$ (C) and $\SI{1.61}{\milli \volt}$ (D) corresponding to the energies $\mp E_{\beta}$ of the $h$ ($\beta^-$) and $p$ ($\beta^+$) components, respectively, of the $\beta$ YSR state
      ($\vstab=\SI{5}{\milli \volt}$,
      $\istab=\SI{1}{\nano \ampere}$, $\vmod = \SI{100}{\micro \volt}$). Scale bar in (C), $\SI{0.3}{\nano \meter}$.
      }
 \end{figure}

We start with the investigation of the native YSR states of single Fe atoms on the surface of a Ag island of thickness $\SI{12}{\nano \metre}$  grown on the (110) surface of a SC Nb single crystal~\cite{Tomanic2016,Schneider2023} (for experimental procedures, see the Materials and Methods section in Supplementary Note 1~\cite{supplementary}). To, first, exclude the interaction of the Fe $3d$ orbitals with any surface-state related mode, we place the Fe atoms inside a nearly square-shaped double-walled quantum corral made out of nonmagnetic (Supplementary Note 2) Ag atoms with a sufficiently small length/width of $L_x=\SI{5.53}{\nano \metre}$ / $L_y=\SI{5.98}{\nano \metre}$ (Fig.~\ref{fig1}A) such that the lowest quantum corral eigenmode has an energy far above $E_{\rm F}$ (Supplementary Note 4 and~\cite{Schneider2023}). Thus, the energy, spatial extension, and $p$-$h$ composition of the native YSR state stemming from the coupling of the Fe $3d$ orbitals to the proximitity-SC Ag bulk states can be determined. Figure~\ref{fig1}B (black curve) shows a Ag substrate spectrum measured in an even smaller empty corral (fig. S2) revealing the de Gennes -- Saint James coherence peaks of the Ag island.
Note, that we used a SC tip with a gap $\Delta_{\rm t}=\SI{1.32}{\milli \eV}$ for all measurements, such that sample states appear at a bias voltage $V$ shifted by $\pm \Delta_{\rm t}/e$ away from zero bias with respect to their original energy (Supplementary Note 1 and fig. S1). 
In comparison to the spectrum taken on the Ag substrate, the spectrum taken on the Fe atom (red curve in Fig.~\ref{fig1}B) reveals three pairs of in-gap peaks (Supplementary Note 2 and fig.~S3) 
which we identify as YSR states resulting from the coupling of the Fe $3d$ orbitals to the Ag island~\cite{Ruby2016}. In the following, we focus on the lowest energy YSR state labeled $\beta^\pm$ (see Fig.~\ref{fig1}B) with energy 
$\pm E_{\beta}$
which is energetically well-separated from the de Gennes -- Saint James coherence states of the substrate, as well as from the MSSs for any corral size.
Constant-contour $\didv$ maps taken at the bias voltage of the $\beta^\pm$ YSR state reveal the spatial shape, extent and $p$-$h$  
composition of this native YSR state (Figs.~\ref{fig1}C,D). 
Its $h$-like ($\beta^-$) and $p$-like ($\beta^+$) components possess different spatial forms, resembling downward and upward pointing triangles, respectively, reflecting an orbital origin that is distinctly different from that of the other two YSR states (Supplementary Note 2 and fig.~S3). Moreover, they have a spatial extent and hence localization length $\approx\SI{0.75}{\nano \metre}$ (see figs. S3G,H), which is similar to the apparent diameter of the Fe atom extracted from the STM images. Finally, the intensity of the $\beta^-$ component is significantly larger than that of the $\beta^+$ component.
We thus conclude that, without the coupling to a surface-related mode, the native $\beta$ YSR state of the Fe atom, which originates in the coupling of the Fe $3d$ orbitals to the paired electrons in the bulk of the Ag island, possesses a dominant $h$-like component and is detectable only up to distances of about $\SI{1}{\nano \metre}$ away from the Fe atom. In the following, we show that both of these features can be significantly altered when the YSR state couples to a spatially extended corral eigenmode tuned close to $E_{\rm F}$.



\section*{YSR quantum projection in elliptical quantum corrals}

\begin{figure}[H]
\centering
	\includegraphics[width=\textwidth]{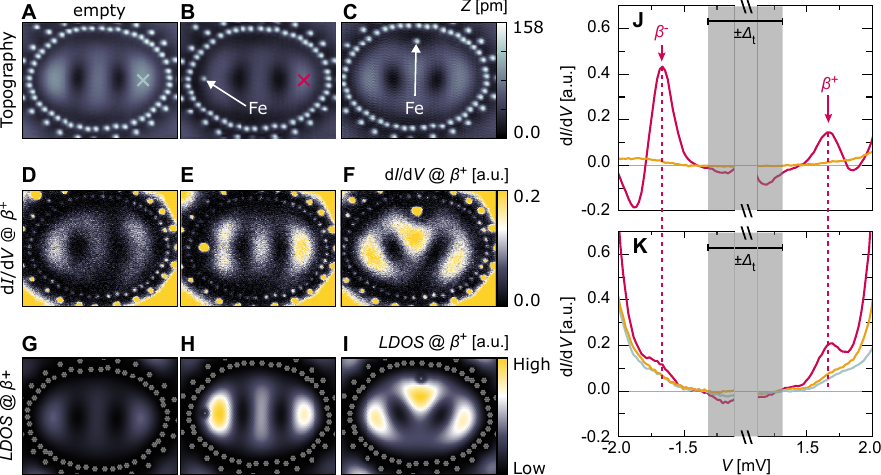} 
 	\caption{\label{fig4} \textbf{YSR quantum projection in an elliptical quantum corral.} (\textbf{A} to \textbf{C}) Constant-current STM images of an empty Ag corral (A) with major axis length $a=\SI{18.20}{\nano \meter}$ and minor axis length $b=\SI{13.30}{\nano \meter}$
  and the same corral with an Fe atom located on the major axis close to the left edge (B) and on the minor axis close to the top edge (C) ($\vbias=\SI{-5}{\milli \volt}$, $\iset = \SI{1}{\nano \ampere}$).
  (\textbf{D} to \textbf{F} and \textbf{G} to \textbf{I}) Experimental (D to F) constant-height $\didv$ maps and simulated (G to I, see Supplementary Note 3) LDOS maps of the empty corral (D, G), the corral with the Fe on the major (E, H) and on the minor axis (F, I) taken close to $+E_{\beta}$ (D to F: $\vbias = \SI{1.67}{\milli \volt}$, $\vstab = \SI{-5}{\milli \volt}$, $\istab=\SI{1}{\nano \ampere}$, $\vmod=\SI{100}{\micro \volt}$; G to I: for simulation parameters, see Supplementary Note 3). (D) to (F) have been set to the same ranges in the color scale. The same applies to (G) and (H). The color scale range in (I) was set to its min. and max. values.
  (\textbf{J}) $\didv$ spectra taken on the Fe atom in (B) (red) and on a Ag atom replacing the Fe atom (orange, see corral in figs.~S4J,M,N). (\textbf{K}) $\didv$ spectra taken on the empty locations indicated by the crosses in (A,B) of the empty corral (gray), of the corral in (B) with the Fe atom (red) and of the corral where the Ag atom replaces the Fe atom (orange) ($\vstab = \SI{-5}{\milli \volt}$, $\istab=\SI{1}{\nano \ampere}$, $\vmod=\SI{50}{\micro \volt}$).
  }
 \end{figure}

To realize the YSR quantum projection, we first consider an elliptical quantum corral whose wall also consists of two rows of nonmagnetic Ag atoms (Fig.~\ref{fig4}A).
The size of the corral is tailored such that there exists a corral eigenmode close to $E_{\rm F}$ 
whose spatial structure is revealed by the close-to-zero-bias  STM image (Fig.~\ref{fig4}A) and constant-height $\didv$ map  (Fig.~\ref{fig4}D) across the corral. 
Placing an Fe atom inside the corral on the main axis close to the left maximum of the corral eigenmode (Fig.~\ref{fig4}B), we find that a $\beta$ YSR state appears in the $dI/dV$ spectrum measured at the Fe site at the same energies $\pm E_{\beta}$ and with similar relative intensities between the 
$\beta^-$ and $\beta^+$
components as in the native case investigated above (Fig.~\ref{fig4}J, cf. Fig.~\ref{fig1}B). However, the spectroscopic signature of this YSR state is now detectable in the $dI/dV$ signal at distances more than ten times larger than the spatial extent of the native $\beta$ YSR state as revealed by a $\didv$ map close to $+E_\beta$ (Fig.~\ref{fig4}E, data close to $-E_\beta$ in fig.~S4E). The spatial pattern of this state strongly resembles that of the empty corral's eigenmode in Fig.~\ref{fig4}D, but it now appears with an increased intensity in Fig.~\ref{fig4}E. 
Moreover, a $\didv$ spectrum taken near the rightmost maximum of this spatially extended state (see red cross in Fig.~\ref{fig4}B) reveals that it is located at approximately the same energy $\pm E_\beta$ and possesses a similar spectral width as the Fe atom's native $\beta$ YSR state (red spectrum in Fig.~\ref{fig4}K). 
Note, however that the $p$-$h$ composition of this extended state is inverted with regards to the native $\beta$ YSR state, with the $p$-like component now showing a larger intensity than the $h$-like component. 
If the Fe atom is replaced by a nonmagnetic Ag atom at the same lattice site (fig.~S4J), the $\didv$ spectrum taken at the position of the cross in Fig.~\ref{fig4}B is largely indistinguishable from that taken on the same site in the empty corral, revealing no signature of the spatially extended in-gap state (orange and gray lines in Fig.~\ref{fig4}K).
The $\didv$ map in Fig.~\ref{fig4}E thus demonstrates that the proximity of the corral's eigenmode to $E_{\rm F}$ leads to a spatially extended projection of the native Fe $\beta$ YSR state, thus creating the theoretically predicted YSR quantum mirage~\cite{Morr2004}. 
Additional data with the Fe atom on other locations inside the corral (Figs.~\ref{fig4}C,F and figs.~S4B,F,G) reveal that the intensity and spatial form of this induced YSR quantum projection strongly depend on the location of the Fe atom with respect to the maxima in the corral eigenmode, 
as expected for the projection induced by a quantum state \cite{Morr2004}. 
Moreover, the nearly identical energies of the $\beta^\pm$ YSR peaks measured on the native Fe and in the presence of the corral eigenmode (cf. Figs.~\ref{fig1}B and ~\ref{fig4}J, see also Supplementary Note 7)
suggest a dominant coupling of the Fe $3d$ orbitals to the bulk Ag electronic states, rather than to the Ag surface state, as the latter would lead to a substantial variation of the YSR state's energy when the size of the corral is changed~\cite{Morr2004}.


To further elucidate the microscopic origin of the experimental data, we consider a semi-infinite 3D tight-binding Hamiltonian $H=H_{\rm 3D} + H_{\rm s} + H_{\rm at} + H_{\rm c} + H_{\rm hyb}$ (for details see Supplementary Note 3), where $H_{\rm 3D}$ describes the semi-infinite 3D electronic structure of the Ag island, $H_{\rm s}$ describes the electronic Shockley surface-state band, $H_{\rm hyb}$ represents the coupling between the surface and bulk states which leads to  proximity induced superconductivity in the former, $H_{\rm c}$ describes coupling of a corral of nonmagnetic Ag atoms to the surface states, giving rise to the eigenmode structure of the surface band inside the corral, and $H_{\rm at}$ represents the scattering arising from the presence of a Ag or Fe atom inside the corral, where we 
assume a dominant coupling of a single orbital on the Fe (Ag) atoms to the bulk (surface) states, as discussed above. The theoretically computed LDOS (Figs.~\ref{fig4}G to I) obtained from this Hamiltonian shows good agreement with the experimental \didu maps (Figs.~\ref{fig4}D to F) for an empty corral, as well as for the cases where an Fe atom is located at different positions inside the corral. 
This strongly substantiates our conclusion that the experimental data can be interpreted as the coherent projection of the native $\beta$ YSR state through the corral's eigenmode, thus creating a YSR mirage. This projection occurs via an indirect path involving a dominant coupling of the Fe $3d$ orbitals to the Ag bulk states which in turn are coupled to the surface state, as schematically shown in Fig.~\ref{fig0}.
We next investigate whether the long-range projected image preserves the nature of the native YSR state coherently.



\section*{Coherent nature of the YSR quantum projection}

\begin{figure}[H]
\centering	
 \includegraphics[width=\textwidth]{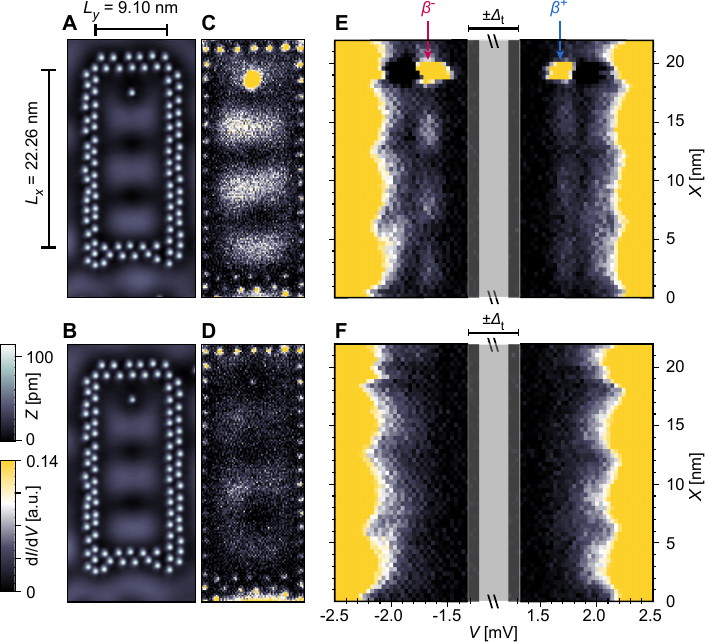} 
 	\caption{\label{fig2} \textbf{Long-range coherence of the YSR quantum projection.} (\textbf{A} and \textbf{B}) Constant-current STM images of a Ag corral ($L_x = \SI{22.26}{\nano \metre}$, $L_y = \SI{9.1}{\nano \metre}$) with an Fe atom placed in the topmost quarter (A) and of the same corral where the Fe atom was replaced with a Ag atom (B) (Fourier-filtered, $\vbias = \SI{-5}{\milli \volt}$, $\iset =\SI{1}{\nano \ampere}$).
  (\textbf{C} and \textbf{D}) Constant-height $\didv$-maps taken close to $-E_{\beta}$ inside the corrals of (A) and (B), respectively ($\vbias=\SI{-1.67}{\milli \volt}$, $\vstab=\SI{-5}{\milli \volt}$, $\istab = \SI{1}{\nano \ampere}$, $\vmod = \SI{100}{\micro \volt}$).
   (\textbf{E} and \textbf{F}) $\didv$ line profiles taken along the longitudinal vertical axes through the corrals in (A) and (B), respectively ($\vstab = \SI{-5}{\milli \volt}$, $\istab =\SI{1}{\nano \ampere}$, $\vmod = \SI{50}{\micro \volt}$).
   The red and blue arrows above (E) indicate $\vbias$ corresponding to $-E_{\beta}$ and $+E_{\beta}$, respectively. The gray vertical lines indicate the bias voltage of the tip gap $\Delta_{\rm t}/e$.}
 \end{figure}

To demonstrate that a coherent projection
of the YSR state can be achieved, we consider a rectangular quantum corral of width $L_y=\SI{9.1}{\nano \metre}$ and length $L_x=22.26$ with an Fe atom  placed on the corral's longitudinal axis (Fig.~\ref{fig2}A). For such a corral, the eigenmodes are
characterized by a pair of quantum numbers $(n_x, n_y)$  reflecting the numbers of maxima $n_x$ ($n_y$) in the $x$- ($y$-) direction~\cite{Schneider2023}. For $L_x=22.26$, the $(3,1)$ eigenmode is the one located closest to $E_{\rm F}$, as evidenced by the constant-current map shown in (Fig.~\ref{fig2}A)  (note a slight downward shift of the eigenmode due to the presence of the magnetic impurity). The native YSR state of the Fe atom couples to this eigenmode, leading to its spatial projection, as follows from the $dI/dV$ map taken close to $-E_\beta$ (Fig.~\ref{fig2}C). A $\didv$ line profile taken along the corral's longitudinal axis (see Fig.~\ref{fig2}E and extended data in Supplementary Note 5 and fig.~S16) shows a clear energetical separation of the $\beta^\pm$ components of the YSR quantum projection from the MSSs, with an energy linewidth comparable to that of the native $\beta$ YSR state.
It also reveals a constant relative phase shift in the spatial oscillations between the $\beta^-$ and $\beta^+$ components of the YSR quantum projection inherited from the MSS (see Supplementary Note 6 and fig.~S18), which is a characteristic signature of its phase coherence~\cite{Balatsky2006,kim2020long,Menard2015}. In addition, we find that the amplitudes of the spatial oscillations hardly attenuate with distance from the Fe atom, which further substantiates the coherence of the projection of the quantum state.
We, thus, conclude that the coupling of the YSR state to the quantum corral eigenmodes leads to its coherent projection, 
which is the fundamental requirement for the non-local detection of quantum states.  Our interpretation of Figs.~\ref{fig2}C and E as showing a projection of the native YSR state is further supported by the observation that, when the Fe atom is replaced by a nonmagnetic Ag atom, no in-gap state exists at $\pm E_\beta$ (see Figs.~\ref{fig2}B,D,F). 
Finally, the YSR quantum projection in this particular rectangular corral, in contrast to the elliptical one, possesses the same qualitative $p$-$h$ composition as the native Fe YSR state (cf. Figs.~\ref{fig2}E,~\ref{fig4}K and~\ref{fig1}B). This raises the intriguing question of whether this composition, and not only the spatial structure of the projection, can be manipulated by shifting the energies of corral eigenmodes through $E_{\rm F}$.

\section*{Manipulating the spatial form and $p$-$h$ composition of projected YSR states}

\begin{figure}[H]
\centering	
 \includegraphics[width=\textwidth]{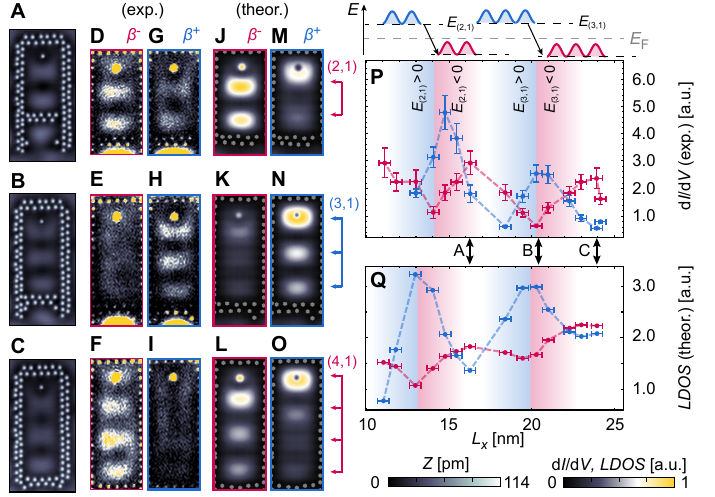} 
 	\caption{\label{fig3} \textbf{Oscillation of particle-hole composition of the YSR quantum projection.} (\textbf{A} to \textbf{C}) Constant-current STM images of Ag corrals, including an Fe atom at the top, with constant widths $L_y =\SI{9.1}{\nano \metre}$ and different lengths $L_x =\SI{16.29}{\nano \metre}$ (A), $\SI{20.27}{\nano \metre}$ (B), and $\SI{23.91}{\nano \metre}$ (C) ($\vbias =\SI{-5}{\milli \volt}$, $\iset = \SI{1}{\nano \ampere}$). 
  (\textbf{D} to \textbf{I}) Experimental constant-height $\didv$ maps and (\textbf{J} to \textbf{O}) simulated LDOS maps taken inside the corrals of (A to C) close to $-E_{\beta}$ (D to F: $\vbias=\SI{-1.67}{\milli \volt}$ and J to L: for simulation parameters, see Supplementary Note 3) and $+E_{\beta}$ (G to I: $\vbias=\SI{1.67}{\milli \volt}$ and M to O: for simulation parameters, see Supplementary Note 3) ($\vstab=\SI{-5}{\milli \volt}$, $\istab = \SI{1}{\nano \ampere}$, $\vmod = \SI{100}{\micro \volt}$).
(\textbf{P}) Experimental and (\textbf{Q}) simulated intensities of the $\beta^-$ (red) and $\beta^+$ (blue) components of the YSR quantum projection as a function of corral lengths $L_x$ extracted from corrals of lenghts $L_x =\SI{4.7}{\nano \metre}$ to $L_x =\SI{24.1}{\nano \metre}$ as described in the text.
The sketch on the top of (P) and corresponding blue and red shaded areas indicate the lengths where the energies $E_{(n_x,n_y)}$ of the eigenmode with the given quantum number $(n_x,n_y)$ cross $E_{\rm F}$ (Supplementary Notes 3 and 4, figs.~S9 and S14).
The double arrows underneath panel P indicate the lengths of the corrals shown in (A to O).}
 \end{figure}

To investigate this question, we consider a series of quantum corrals with increasing length $L_x$, allowing us to shift the energies of several eigenmodes through the SC gap.
Figures~\ref{fig3}A-C show constant-current STM images for three different corral lengths, with the corral eigenmodes being located at different distances from $E_{\rm F}$. The corresponding $\didv$ maps taken inside the corrals close to $-E_\beta$ and $+E_\beta$ shown in Figs.~\ref{fig3}D to F and Figs.~\ref{fig3}G to I, respectively, reveal again the existence of a YSR quantum projection; 
they are well reproduced by the corresponding theoretically computed LDOS shown in Figs.~\ref{fig3}J to L, and Figs.~\ref{fig3}M to O, respectively. While for the corrals shown in Figs.~\ref{fig3}A and C, with eigenmodes located well outside the gap (for details, see Supplementary Notes 3 and 4) the intensity of the $\beta^-$ component is larger than that of the $\beta^+$ component, this $p$-$h$ composition is reversed for the corral shown in Fig.~\ref{fig3}B, where the $(3,1)$ corral eigenmode in the absence of SC would be located very close to $E_{\rm F}$. A plot of the $\beta^-$ and $\beta^+$ quantum projection intensities for a larger number of quantum corral lengths $L_x$ (see fig.~S17 and Supplementary Note 6) reveals an oscillatory behavior of the intensities, with the $\beta^+$ projection intensity always being larger than the $\beta^-$ projection intensity whenever a corral eigenmode is very close to $E_{\rm F}$ (see sketch on top of Fig.~\ref{fig3}P, the blue/red shaded areas in Fig.~\ref{fig3}P show those ranges where the respective corral eigenmode is above/below $E_{\rm F}$). At the same time, the energy of the YSR projection remains essentially unchanged and close to $\pm E_\beta$ (see figs.~S20A, S21). These features are well reproduced by the theoretically computed intensities shown in Fig.~\ref{fig3}Q (see Supplementary Note 3 for details). Thus, by tuning the energy of a corral eigenmode close to $E_{\rm F}$, it is possible to tune the $p$-$h$-composition of the YSR projection, as reflected in the intensities of the $\beta^+$ and $\beta^-$ components. Moreover, this manipulation of the $p$-$h$-composition can occur while the projections' spatial shape remains essentially unaltered, as follows from a comparison of the $\didv$ maps for corrals of length $L_x=\SI{18.4}{\nano \metre}$ and $\SI{20.3}{\nano \metre}$ (fig.~S17). 
%

The dependence of the YSR projection on the corral length (see Figs.~\ref{fig3}P and Q), and hence on the proximity of the eigenmodes to $E_{\rm F}$, is not universal. Indeed, we find in the model that the qualitative nature of this dependence varies greatly with the strength of the hyridization $V_{\rm hyb}$ between the Shockley surface-states and the Ag island bulk bands (see fig.~S10 in Supplementary Note 3). We ascribe these qualitative differences to the varying $p$-$h$ asymmetry (with changing $V_{\rm hyb}$ and corral length) that is imposed on the entire system by the highly $p$-$h$ asymmetric corral eigenmodes.  Indeed, by reducing the complexity of the  theoretical model to a model in which a single orbital of an impurity is coupled to a simplified zero-dimensional but still highly $p$-$h$ asymmetric MSS via the Ag bulk states, we reproduce not only the qualitative behavior shown in Figs.~\ref{fig3}P and Q (see fig.~S22 in Supplementary Note 8), but also the dependence on $V_{\rm hyb}$ (fig.~S23). Interestingly, the relative intensities of the $\beta^\pm$ components abruptly change when the YSR state crosses $E_{\rm F}$ and the spin ground state of the SC changes (see fig.~S12 in Supplementary Note 3 and fig.~S22 in Supplementary Note 8)~\cite{Balatsky2006}. Both models thus reveal that the coupling of the native YSR state in the bulk Ag island to the surface corral eigenmodes allows for great flexibility in manipulating the $p$-$h$ composition of the coherent YSR projection. 

\vspace{-10pt}
\section*{Conclusions}
\vspace{-10pt}
In summary, we experimentally realized a hybrid quantum state between a local bulk YSR state and delocalized proximity-induced superconducting quantum corral eigenmodes, enabling a coherent projection of the former over length scales 20 times larger than its localization length and providing us with a minimal-invasive non-local capability to probe the quantum state of this single atom magnet-SC hybrid system via STM. Furthermore, we demonstrated that the $p$-$h$ composition of the YSR projection can be reversed by tuning a corral eigenmode close to $E_{\rm F}$, which can be achieved by changing the shape and size of the quantum corral. The predicted inversion of the $p$-$h$ composition when the native YSR state is shifted across $E_\textrm{F}$ could help to identify the quantum phases of single-atom, multi-orbital YSR states on SC surfaces~\cite{Liebhaber2020}. We foresee intriguing possibilities for future work if the Fe was replaced by magnetic atoms that couple more strongly to the Shockley surface-state~\cite{Moro-Lagares2018, Aapro2023}, thereby realizing more strongly coupled YSR quantum corral hybrid states. In such systems, the feasibility of engineering the strength and landscape of couplings of distant atomic YSR states was theoretically predicted~\cite{Morr2004,Chiappe2002,Correa2002,Stepanyuk2005,Stepanyuk2007,Brovko2009,Ngo2017,Hallberg2002}. Finally, the range and potential of the above demonstrated methodology might present a new approach to non-locally access and manipulate YSR qubits~\cite{mishra2021YuShibaRusinovQubit} or elusive quantum states like $p$-$h$ symmetric Majorana bound states~\cite{Schneider2022,Gharavi2016,Steiner2020}.

\newpage

\bibliographystyle{Science}

\newpage
\section*{Acknowledgments}
L.S., T.P., I.I., J.W. and R.W. gratefully acknowledge funding by the Cluster of Excellence ‘Advanced Imaging of Matter’ (EXC 2056 – project ID 390715994) of the Deutsche Forschungsgemeinschaft (DFG). K.T.T. and R.W. acknowledge funding of the European Union through the ERC Advanced Grant ADMIRE (project no. 786020).
J.W. acknowledges support by the DFG – project WI 3097/4-1 (project No. 543483081). C.X.\ and D.K.M.\ acknowledge support by the U.\ S.\ Department of Energy, Office of Science, Basic Energy Sciences, under Award No.\ DE-FG02-05ER46225.  T.P. acknowledges funding from the European Union (ERC, QUANTWIST, project number 101039098).

\section*{Supplementary materials}
Supplementary Note 1, Materials and Methods\\
Supplementary Notes 2 to 8\\
figs. S1 to S23\\
References \textit{(1-12)}


\clearpage


\end{document}